\documentclass[english]{article}
\usepackage[T1]{fontenc}
\usepackage[latin9]{inputenc}
\usepackage{float}
\usepackage{amssymb}
\usepackage{graphicx}

\makeatletter

\providecommand{\tabularnewline}{\\}

\newcommand{\lyxaddress}[1]{
\par {\raggedright #1
\vspace{1.4em}
\noindent\par}
}
\usepackage{epsfig}
\usepackage{lscape}
\date{}
\usepackage{url}

\@ifundefined{showcaptionsetup}{}{%
 \PassOptionsToPackage{caption=false}{subfig}}
\usepackage{subfig}
\makeatother

\usepackage{babel}
\begin{document}

\title{Beam Charge Measurement for the g2p/GEp experiments}

\author{Pengjia Zhu}
\maketitle
\lyxaddress{\begin{center}
University of Science and Technology of China
\par\end{center}}
\begin{abstract}
The g2p/GEp experiments used a solid $\mathrm{NH_{3}}$ polarized target, where the polarization of the target is sensitive to temperature and radiation. The beam current was limited to 5-100 nA during the experiment to avoid too much depolarization of target (The typical Hall A running condition for beam current is 1 $\mu A$ to 100 $\mu A$). The measured charge was further used to get the accurate physics cross sections. New BCM (Beam Current Monitor) receivers and a DAQ system were used to measure the beam current at such a low current range. A tungsten calorimeter was used to calibrate the BCMs. This technical note summarizes the calibration procedure and the performance of the BCMs.
\end{abstract}

\section{Setup}

\begin{figure}
\begin{centering}
\includegraphics[width=1\linewidth]{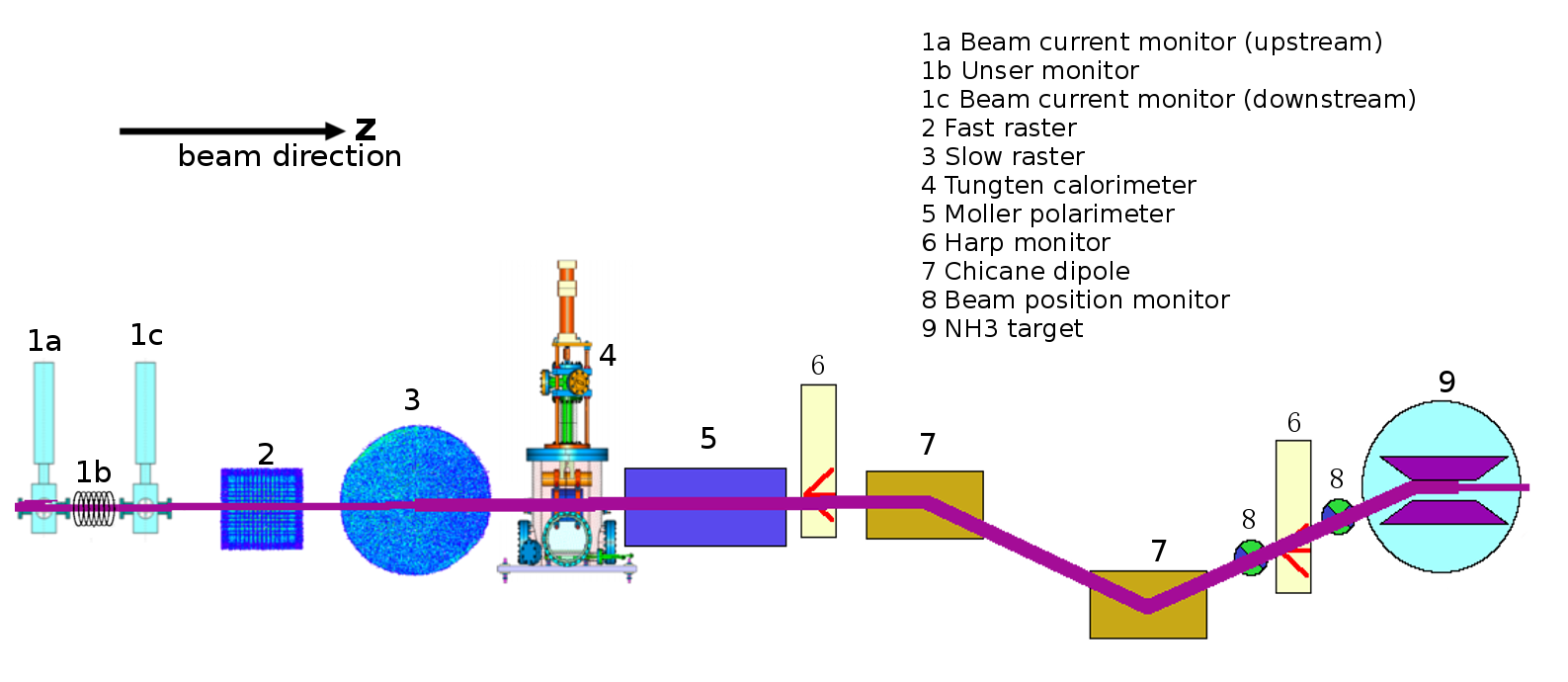}
\par\end{centering}

\caption{\label{fig:Beamline}Beamline for the g2p/GEp experiments}

\end{figure}

The BCM system used for the g2p \cite{g2pproposal} and GEp \cite{gepproposal} experiments contains two RF cavities (1a and 1c in Fig. \ref{fig:Beamline}), BCM receivers with associated data-acquisition (DAQ) system, and a tungsten calorimeter for calibration. The traditional calibration method using an Unser monitor \cite{Unser} (1b in Fig. \ref{fig:Beamline}) would not work at low current because it has an accuracy of 0.3 $\mu A$. Using a Farady cup \cite{Faradaycup} in the injector region to calibrate BCMs there and then using the calibrated BCMs in the injector region to cross-calibrate the BCMs in Hall A has about the same accuracy due to the beam loss in the machine. The setup is shown in Fig. \ref{fig:Beamline}.

\subsection{BCM receiver}

\begin{figure}
\begin{centering}
\includegraphics[width=1\linewidth]{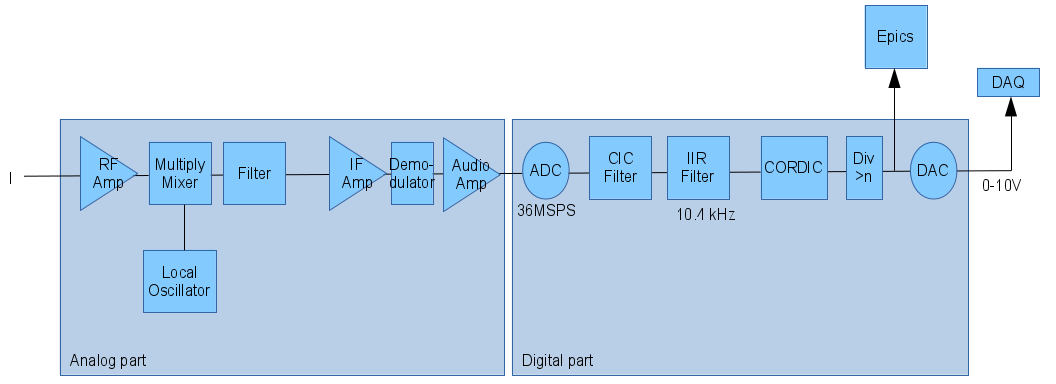}
\par\end{centering}

\caption{\label{fig:BPM-receiver}BCM receiver used for the g2p/GEp experiments}
\end{figure}

Since the original RMS-to-DC converter \cite{Denard2001} did not work at low current, new BCM receivers were designed by John Musson and his colleagues from the JLab instrumentation group for the purpose of achieving a reasonable signal/noise (S/N) ratio in the beam current range of several nanoampere to several micro-ampere \cite{mussonece652paper}. The design diagram is shown in Fig. \ref{fig:BPM-receiver}.

The receiver consists of an analog part and a digital part. The analog part includes the amplifier and the mixer. The multiply mixer converts the ratio frequency (RF) signal to the intermediate frequency (IF) signal. The signal is digitized by a 36 MSPS ADC, and applied by a cascaded-integrator\textendash comb (CIC) filter and an infinite-inpulse-response (IIR) filter (10.4 kHz). The CORDIC system is used to get the amplitude and phase of the digital signal \cite{mussonece652paper}. The 20-bit digital signal is converted back to 0-10V analog signal to match the existing Hall A DAQ system using a 18-bit DAC. A DIV unit is used to intercept the signal from 20-bit to 18-bit by applying an adjustable bit shift. More details can be found in \cite{mussonece652paper}.

\subsection{Data acquisition system}

\begin{figure}
\begin{centering}
\includegraphics[width=0.8\linewidth]{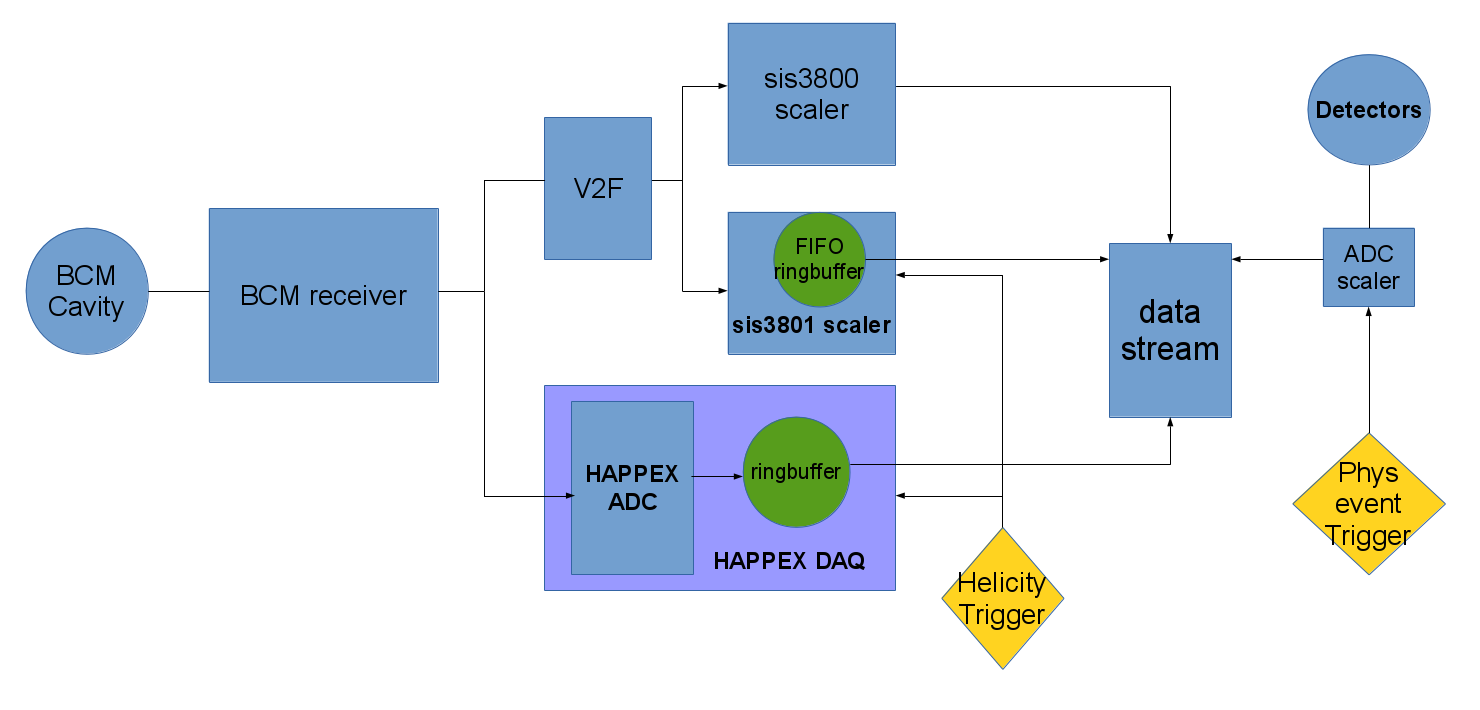}
\par\end{centering}

\caption{\label{fig:daq_bcm}DAQ system for BCM}

\end{figure}

The BCM data from the receivers were passed on the DAQ system as shown in Fig. \ref{fig:daq_bcm}.
 The voltage signal from receiver was split and sent to the Voltage to Frequency (V2F) module and the HAPPEX ADC.

\subsubsection{Helicity}

The beam is polarized in injector before going to the CEBAF accelerator. The polarization is controlled by a helicity control board (VME) \cite{helicityboard}. The helicity control board generates several signals which relative to each other. It controls the high voltage supply to change the orientation of the polarization of laser, which is used to generate the polarized electron beam with GaAs photogun by using the method of optical pumping. Meanwhile the helicity control board sends signals to the DAQ system in the hall in order to get the helicity based information. During the experiment the helicity setting was the same as the QWEAK experiment in Hall C, as shown in table \ref{tab:Helicity-configuration}.
\begin{table}
\begin{centering}
\begin{tabular}{|c|c|}
\hline 
Mode & Free clock\tabularnewline
\hline 
\hline 
T-Settle & 70 $\mu s$\tabularnewline
\hline 
T-Stable & 971.65 $\mu s$\tabularnewline
\hline 
Helicity Pattern & $+--+$ or $-++-$\tabularnewline
\hline 
Reporting delay & 8 window\tabularnewline
\hline 
Helicity board frequency & 960.015 Hz\tabularnewline
\hline 
\end{tabular}
\par\end{centering}

\caption{\label{tab:Helicity-configuration}Helicity configuration}
\end{table}

\begin{figure}
\begin{centering}
\includegraphics[width=0.9\columnwidth]{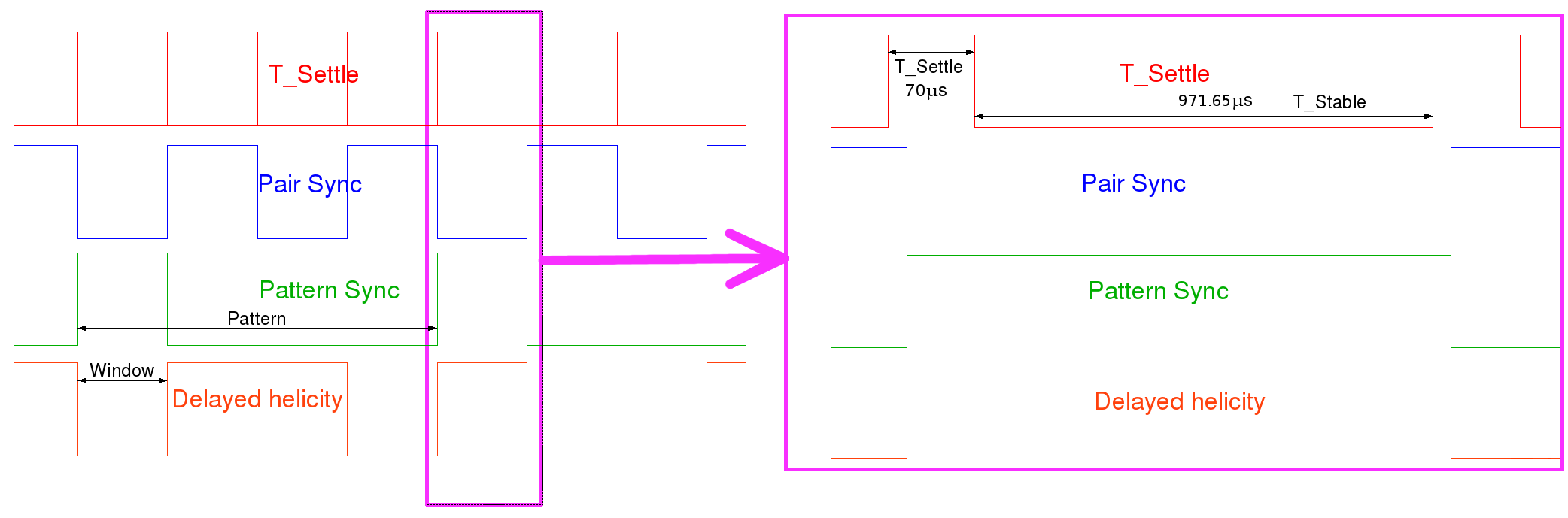}
\par\end{centering}

\caption{\label{fig:helicity}Helicity signal from helicity control board. }
\end{figure}

Four helicity signals were sent to hall during the experiment via optical fibers, which named T-Settle (or MPS), pattern\_sync (or QRT), pair\_sync, and delayed helicity (Fig.  \ref{fig:helicity}). The quartet helicity pattern is used for the experiment to minimize the system error, which is ``$+--+$'' or ``$-++-$'', one pattern is composed with four helicity windows. The pattern\_sync indicates the first window of one pattern. The T-Settle signal is used to indicate if the helicity is valid. The high-level T-Settle (70 $\mu s$) indicates the helicity flips, or has unsure helicity states, while the low-level T-Stable (971.65 $\mu s$) indicates the reliable helicity states. The pair sync signal flips in each helicity window, which is used as the redundancy information. The helicity flip signals sent to the halls are 8 windows delayed with respect to the actual helicity flip signals, and need to be taken care of in decoding. More details about the helicity decoder can be found in \cite{chaohelicity}.

\subsubsection{Scaler}

The V2F converts the DC voltage signal from the BCM receiver to a frequency signal in order to readout the scalers for counting. The SIS380x scaler has two modes selected by a jumper on the board: SIS3800 and SIS3801.

\paragraph{SIS3800 scaler}

The SIS3800 scaler counts the charge, clock and trigger signals for each event, and delivers them to the data stream when the event trigger is accepted. The counter data for the SIS3800 is only cleared at the beginning of the run, thus the SIS3800 is used to get the counts for the whole run.

\paragraph{SIS3801 scaler}

\begin{figure}
\begin{centering}
\includegraphics[width=0.7\columnwidth]{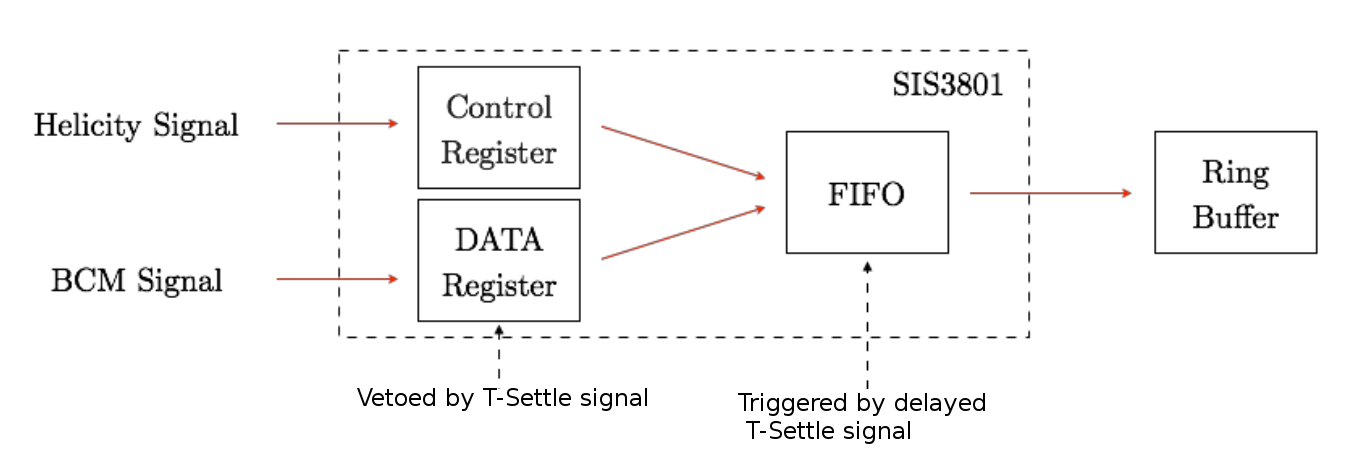}
\par\end{centering}

\caption{\label{fig:sis3801-workflow}Workflow of the SIS3801 scaler \cite{chaohelicity} }
\end{figure}

\begin{figure}[H]
\begin{centering}
\includegraphics[width=0.8\linewidth]{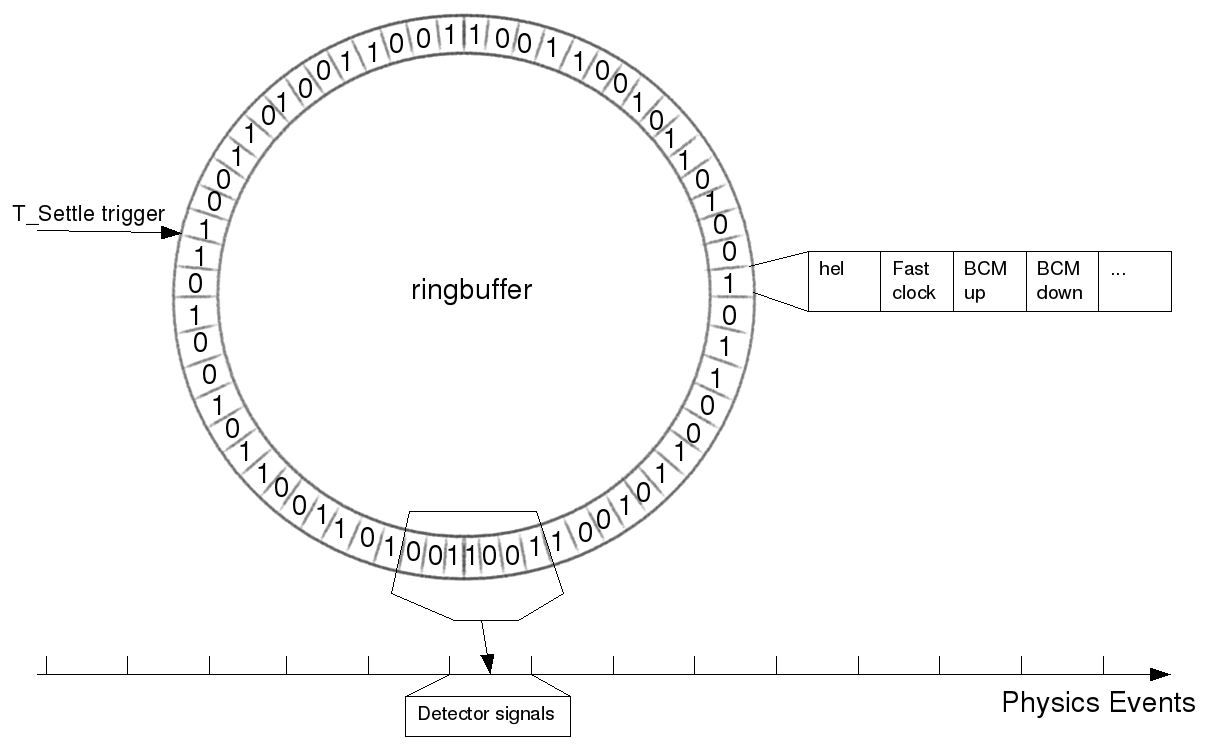}
\par\end{centering}

\caption{\label{fig:Ringbuffer}Workflow of the ringbuffer. The ringbuffer is used as the buffer merging from the data-stream of the helicity triggered DAQ to the physics triggered DAQ. For the SIS3801 scaler, the FIFO register is used as the ringbuffer. For the HAPPEX DAQ, an array defined in the CPU register is used as the ringbuffer.}
\end{figure}

The SIS3801 is used to get the helicity gated information. Fig.  \ref{fig:sis3801-workflow} shows the workflow of the SIS3801 scaler 
. The scaler is controlled by the T-Settle signal. The data registers count the charge, clock and trigger signals only in the T-Stable part of the helicity window. The counts are reset by the high-level T-Settle. A delayed T-Settle, the Pattern Sync, and the delayed helicity are also sent to the control register. Those information are saved in the FIFO (First-In-First-Out) register triggered by the delayed T-Settle signal. The FIFO is used as a ringbuffer (Fig.  \ref{fig:Ringbuffer}) before merging to the standard DAQ system. 

\subsubsection{HAPPEX DAQ}

The HAPPEX DAQ were designed for the parity violation experiments. This DAQ was reprogrammed and reassembled for the g2p/GEp experiments.

\begin{figure}
\begin{centering}
\includegraphics[width=1\columnwidth]{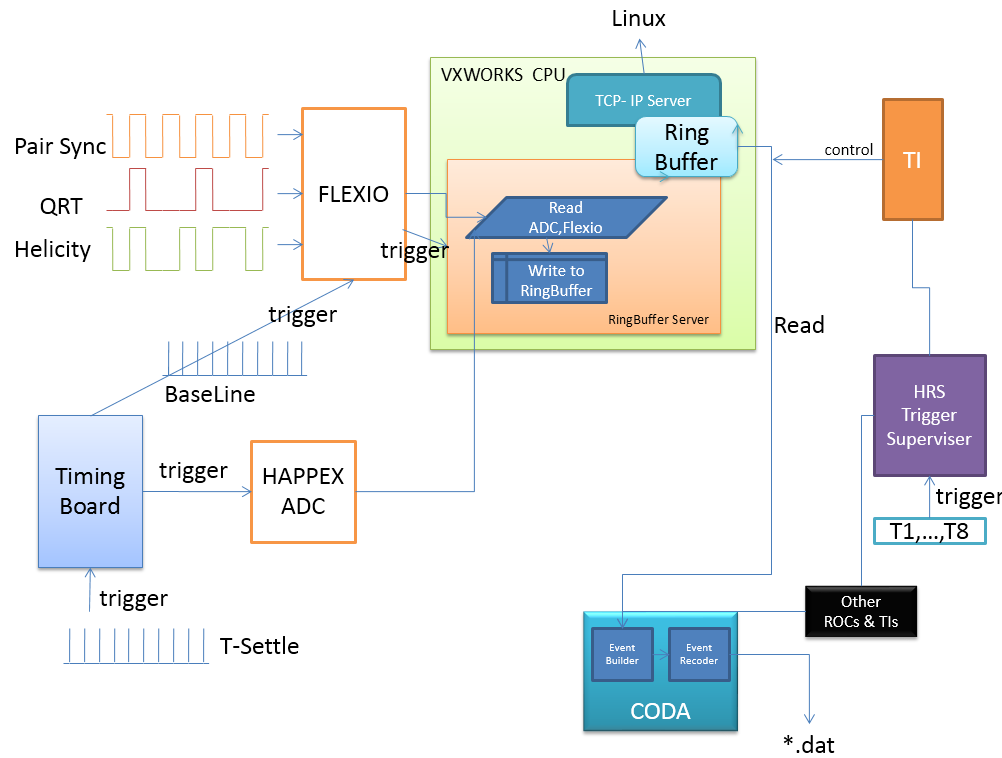}
\par\end{centering}

\caption{\label{fig:Happex-DAQ-diagram-1}HAPPEX DAQ diagram}
\end{figure}

The HAPPEX DAQ contains a timing board (VME) \cite{timingboard}, several 18-bit ADCs \cite{ADC18bob}, a flexible IO (FLEXIO, VME) \cite{flexioEd}, a trigger interface module (TI), and a VxWorks CPU. The diagram of HAPPEX DAQ is shown in Fig. \ref{fig:Happex-DAQ-diagram-1}.

\paragraph{Timing board}

\begin{figure}[H]
\begin{centering}
\includegraphics[width=0.9\linewidth]{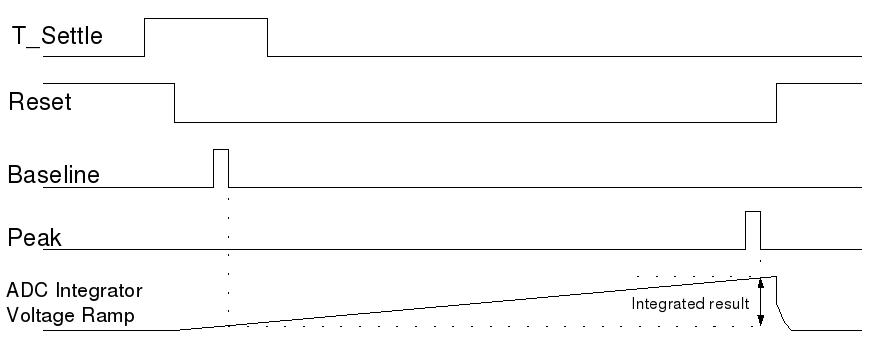}
\par\end{centering}

\caption[Timing board signals]{\label{fig:Timing-board-signals-1}Signals from timing board \cite{flexioEd}}
\end{figure}

The timing board generates several time signals to control the start and stop integration time of the ADCs. The T-Settle signal is used as the trigger source for the timing board. Based on the trigger signal, the timing board generates a set of signals (Fig. \ref{fig:Timing-board-signals-1}). The reset signal controls the ADC integration. The delay time between the baseline signal and the peak signal is used as the integration time, and the digital value difference between them is used as integrated result. The DAC module in the timing board was used as a debugging source during the experiment.

\paragraph{HAPPEX ADC}

\begin{figure}
\begin{centering}
\includegraphics[width=0.95\linewidth]{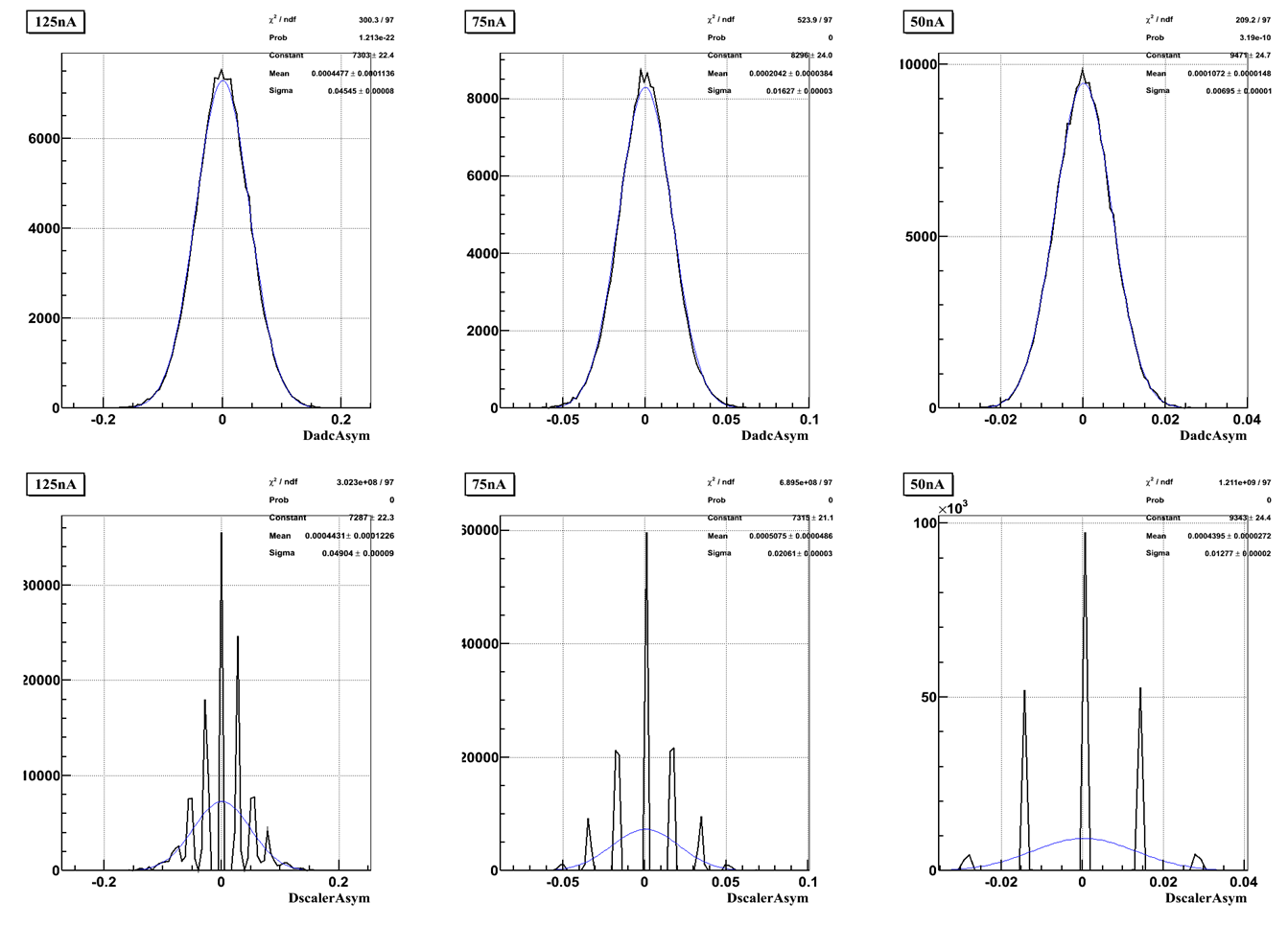}
\par\end{centering}

\caption[Charge asym comparison between HAPPEX and scaler]{\label{fig:Comparison-for-asym}Comparison of charge asymmetry measurements from HAPPEX ADCs and scalers. The top three plots use HAPPEX ADCs, while the bottom three plots use scalers. The beam currents from left to right are 125 nA, 75 nA, and 50 nA. The total number of events are same in each histogram.}
\end{figure}

The HAPPEX ADC is designed for high bit resolution (18-bit) and a small non-linearity ($\leq2\times10^{-5}$) for measuring small parity violating asymmetries to high precision. From the asymmetry measurement test (Fig. \ref{fig:Comparison-for-asym}), the bit resolution for the HAPPEX ADCs were much better than the one for the scalers. The integration time of the HAPPEX ADCs controlled by the timing board is 875 $\mu s$, a little bit smaller than the helicity period (1041.65 $\mu s$). The HAPPEX ADCs record more precise position and current information than the FASTBUS 1881 ADCs (with an integration time less than 50 $ns$ during the experiment). 

\paragraph{Flexible IO}

The flexible IO is used to record the digital information. The baseline signal peak from the timing board triggers the flexible IO to record the helicity signals. It also provides a trigger signal for the ringbuffer.

\paragraph{Ring Buffer}

A VxWorks CPU controls the data reading from the HAPPEX ADCs and the flexible IO to the ringbuffer server in the CPU. The ringbuffer is an array saved in the register of the CPU. Each element in array includes the information of helicity, charge, clock signals for this helicity states (Fig.  \ref{fig:Ringbuffer}). For more reliable performance and less CPU occupation, a trigger is used instead of checking the pair sync polarity all of the time. The trigger from the flexible IO has the same period as the T-Settle. Each trigger causes the CPU to read out the data from the flexible IO and the ADCs once. A trigger interface controlled by the HRS trigger supervisor reads the data from the ringbuffer server to the data-stream. For the online debugging, a TCP-IP server was running on the CPU to readout the data from the ringbuffer from any Linux computer at any time.

\subsection{Tungsten Calorimeter}

\begin{figure}
\begin{centering}
\includegraphics[width=0.2\linewidth]{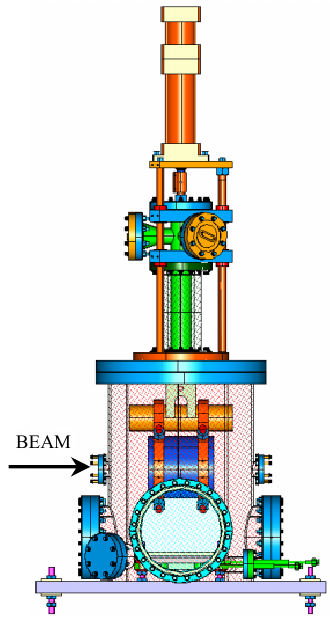}
\par\end{centering}

\caption{\label{fig:Tungsten-Calorimeter}Tungsten Calorimeter}

\end{figure}

A tungsten calorimeter \cite{hacalorimeter} is located downstream of the BCMs and the two rasters \cite{Zhu20161} for calibrating the BCMs by measuring the beam induced temperature rise, as shown in Fig. \ref{fig:Tungsten-Calorimeter}. 
 The chamber that holds the tungsten is pumped down to vacuum to minimize heat loss. The tungsten is in three positions for the different purpose: 
\begin{enumerate}
\item Beam charging, the tungsten is in beam position. All of the incoming beam electrons hit the tungsten. The temperature is increasing during this period.
\item Equilibrating, the tungsten moves out of the beam pipe but doesn't touch the cooling plate. The beam turns off. The temperature stabilizes. The measurement of the temperature occurs in this period.
\item Cooling, the tungsten moves to the cooling plate to cool down the tungsten.
\end{enumerate}
For the temperature measurement, six resistance temperature detectors (RTDs) are mounted on the outer surface at each end of the tungsten slug.

\section{Calibration of the BCMs}

Calibration data were taken several times during the experiment. In order to achieve the uniform heat load from the beam over the tungsten surface, the rasters were turned on during the BCM calibration. The limited size of the ringbuffer caused potential loss of data when the read-out speed was lower than the read-in speed. For the deadtime consideration, the DAQ system only read out no more than 50 sets data from the ringbuffer. An additional clock trigger with a frequency larger than 20 Hz ($\geqq$ 960 (helicity frequency) / 50) was added to avoid data loss in the ringbuffer recorded in the data-stream. The clock signal was needed for calculating the pedestal slope of the scaler and the ADC. For the HAPPEX ADC, the helicity entries were used as the clock. 

The pedestal slope is defined as the accumulated counts or ADC values per unit time when there is no beam. The value of it depends on the frequency of the clock source. It needs to be removed for extracting the real accumulated counts caused by beam. There are two types of clock: fast clock and slow clock. The frequency of the fast clock was $\sim$ 103.7 KHz, while the frequency of the slow clock was $\sim$ 1 KHz. The calibration was taken for each of them. 

\begin{figure}
\begin{centering}
\includegraphics[width=1\linewidth]{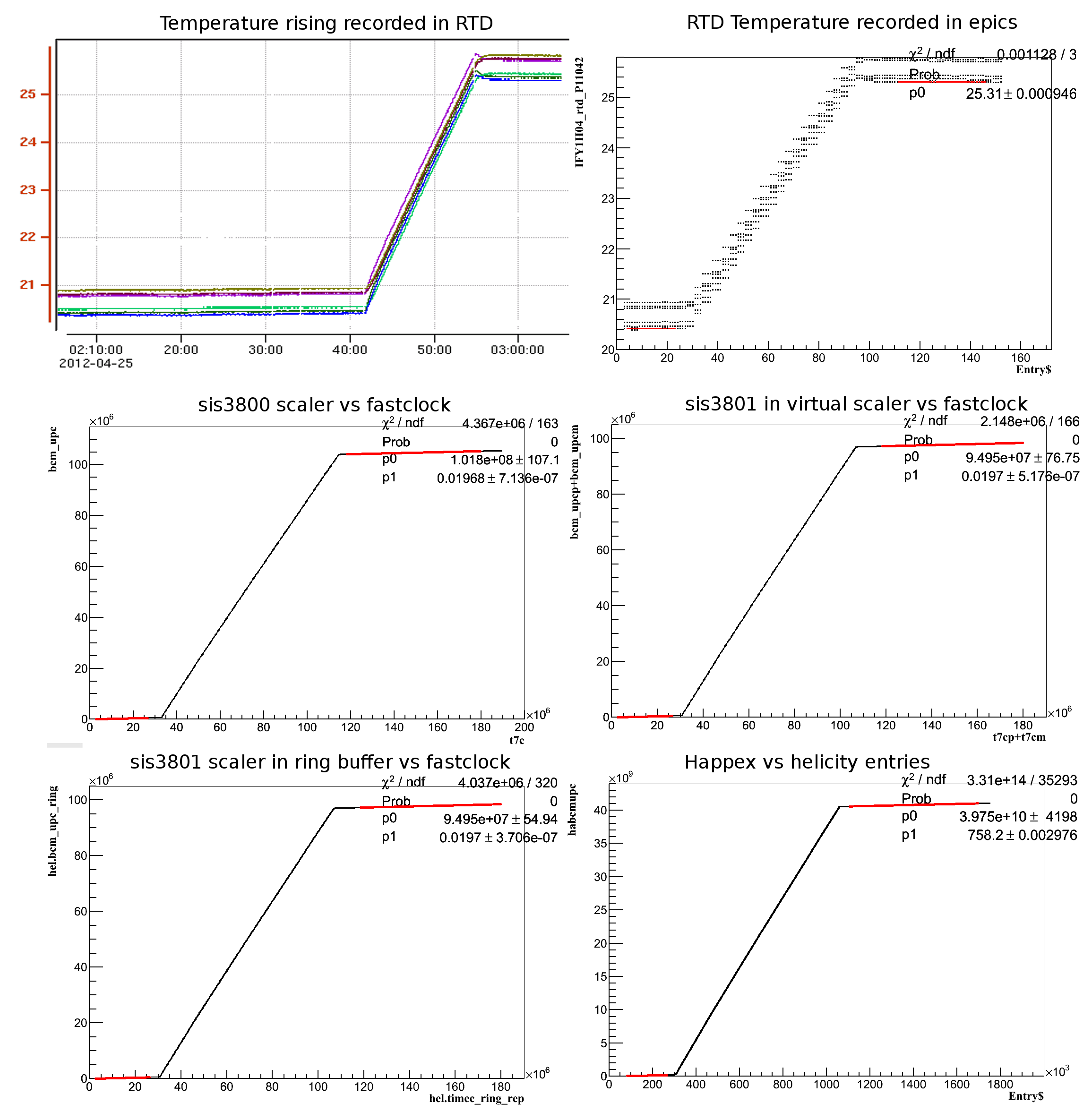}
\par\end{centering}

\caption{\label{fig:BCM-Calibration}BCM Calibration, the top left and right figures are the temperature rise of the RTDs, the last four plots show the counts recorded in the scalers and the HAPPEX ADCs at the same time.}

\end{figure}

\begin{figure}
\begin{centering}
\includegraphics[width=0.7\linewidth]{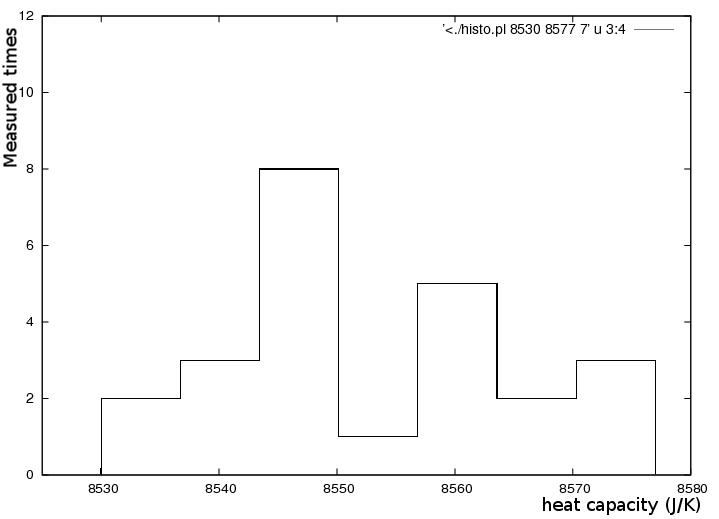}
\par\end{centering}

\caption{\label{fig:Tungsten-calorimeter-heat}Tungsten Calorimeter Heat Capacity Determination \cite{tungstenelog}}
\end{figure}

A complete calibration period is shown in Fig.  \ref{fig:BCM-Calibration}.
 The total temperature rise is used to calculate the total charge. When the beam just off, the temperature readouts keep fluctuating until the heat is uniform in the tungsten. The zero-order polynomial fits are taken before the beam charging and after the temperature become stable when the tungsten is in the equilibrating position. The relationship between the total charge and the temperature rise is:

\begin{equation}
Charge=K\cdot Temperature,\label{eq:c=00003Dkt}
\end{equation}
where $K$ is the heat capacity of tungsten. It was measured by Ahamad Mahmoud before the experiment \cite{tungstenelog}. The result is shown in Fig. \ref{fig:Tungsten-calorimeter-heat}
, with the value of $8555.5\pm50\,J/K$. $Temperature$ is the average temperature from the 6 RTDs. 

There are several devices needed to be calibrated, and each one has its own special condition. The detail calibration procedures for each device are as follows.

\subsection{\label{sub:Calibration-for-SIS3800}Calibration for SIS3800 scaler}

A reset signal was sent to the SIS3800 scaler at the beginning of the run to clear the counts. Since the scaler was found to cause high deadtime, only clock signals were sampled for each event, while other signals were sampled for each 1000 events. Also the DAQ read the scaler once at the end of the run. 

The middle left picture in Fig. \ref{fig:BCM-Calibration} is for the SIS3800 calibration. The rise in the graph is the period when the beam hits the tungsten, corresponding to the rise in the top left. The relation of the total charge and the counts is defined as:

\begin{equation}
Charge=slope\cdot(\Delta counts-pedslope\cdot\Delta clockcounts),\label{eq:sis3800charge}
\end{equation}
where $\Delta counts$ is the total BCM counts accumulated in the scaler, $\Delta clockcounts$ is the total clock counts accumulated in the scaler. The $pedslope$ is the value of the pedestal slope, which is calculated from the first-order polynomial fits before and after the beam. To get the $slope$ value, two time points are chosen before and after the beam heats the tungsten. Using the $\Delta counts$ and the $\Delta clockcounts$ between these two time points and combining with the charge calculated from the temperature, the $slope$ value is then determined.

The beam current is calculated from the calibration constants as: 

\begin{equation}
Current=slope\cdot(rate-pedslope\cdot clockrate),\label{eq:sis3800current}
\end{equation}
where $rate$ and $clockrate$ are defined as the BCM counts per second and clock counts per second.

\subsection{Calibration for SIS3801 scaler}

To calibrate the SIS3801 scaler it is necessary to accumulate all of the counts for each helicity window without any loss of data. There are two methods to get the total counts. One is using the sum counts from two virtual scalers. The offline analyzer \cite{analyzer} automatically accumulates the total counts for positive helicity states and negative helicity states, which present two independent variables (positive and negative virtual scaler) in the raw data. Another is accumulating all of the counts from the ringbuffer. The helicity decoder was used to check if data were lost. The calculated calibration constants are the same from the two methods. The procedures are similar to the SIS3800. The relation of the total charge and the counts is defined as:

\begin{equation}
Charge=slope\cdot(\Delta counts-pedslope\cdot\Delta clockcounts),\label{eq:sis3801charge}
\end{equation}
where the $\Delta counts$ and $\Delta clockcounts$ are counted from the SIS3801 scaler. The value of $Charge$ uses the same value from tungsten as in section \ref{sub:Calibration-for-SIS3800}, thus it is considered as the whole charge in the whole helicity window. Since the SIS3801 does not count for 70 $\mu s$ for each 1041.65 $\mu s$, the slopes calculated for the SIS3801 are larger than the slope for the SIS3800. If we denote $readout$ as the readout from SIS3801 in each helicity entry, the $\Delta clockcounts$ recorded in the SIS3801 for one helicity window is equal to $103700s^{-1}\cdot971.65\mbox{\ensuremath{\mu}s}$, where $103700s^{-1}$ is the frequency of the fast clock, and $971.65\mu s$ is the duration of T-Stable. The beam current is then calculated as the charge divide the duration of the whole helicity window $1041.65\mu s$ :

\begin{equation}
Current=slope\cdot(readout-pedslope\cdot103700s^{-1}\cdot971.65\mbox{\ensuremath{\mu}s})/1041.65\mu s,\label{eq:sis3801current}
\end{equation}
Note the constants in equation (\ref{eq:sis3801charge}) are used to calculate the charge in the whole helicity window. If one need to know the absolute charge in T-Stable, an additional factor of $971.65\mu s/1041.65\mu s$ is needed to be applied.

\subsection{Calibration for HAPPEX ADC}

To calibrate the HAPPEX ADC, the $\Delta counts$ are accumulated for all of the events between two time periods as the total counts. The entries in the HAPPEX DAQ are used as the time stamp. The relation of the total charge and the counts is defined as:

\begin{equation}
Charge=slope\cdot(\Delta counts-pedslope\cdot\Delta entries),\label{eq:happexcharge}
\end{equation}
Similar as the SIS3801 scaler, the ADC only accumulate during the integration time. The beam current is calculated as:

\begin{equation}
Current=slope\cdot(readout-pedslope)/1041.65\mbox{\ensuremath{\mu}s}.\label{eq:happexcurrent}
\end{equation}
Where $readout$ is the readout ADC value. Note the charge calculated using equation (\ref{eq:happexcharge}) is the charge in the whole helicity window.

\subsection{Uncertainty}

The uncertainty of the calculated charge from the tungsten calorimeter comes from the beam energy, RTD, measured tungsten heat capacity, and the heat loss. The ARC measurement has a relative uncertainty of $\sim$ $5\times10^{-4}$ \cite{hallanim}, which contributes to the uncertainty of calculated charge of 2 $nC$ per 1 K temperature rise (for 2.2 GeV beam energy), which is negligible compared to the total charge received in tungsten calorimeter of 30 $\mu C$ during the calibration. The uncertainties of the RTDs are 12.5 mK \cite{calorimeterrtd}, which contributes an uncertainty of $\mathrm{0.046\ \mu C}$. The $50\,J/K$ uncertainty of heat capacity contributes $\mathrm{0.18\ \mu C}$ per 1 K temperature rise. The Hall A calorimeter thermal and mechanical design limits heat losses to $\sim$ 0.2 \% level if the measurement is within 20 min \cite{hacalorimeter}, which causes the uncertainty of calculated charge additional 0.2 \%. The total uncertainty is $\sim$ 0.68 \% for the calculated charge from the tungsten calorimeter.

\begin{figure}
\begin{centering}
\subfloat[Absolute difference between upstream and downstream charge]{\begin{centering}
\includegraphics[width=0.45\linewidth]{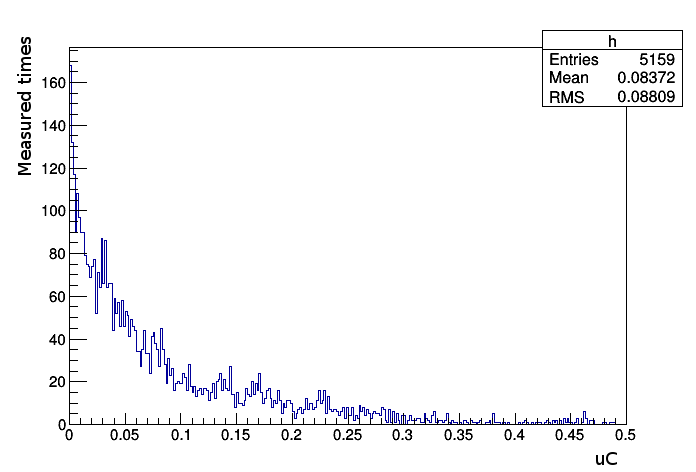}
\par\end{centering}

}$\;$\subfloat[Relative difference between upstream and downstream charge]{\begin{centering}
\includegraphics[width=0.45\linewidth]{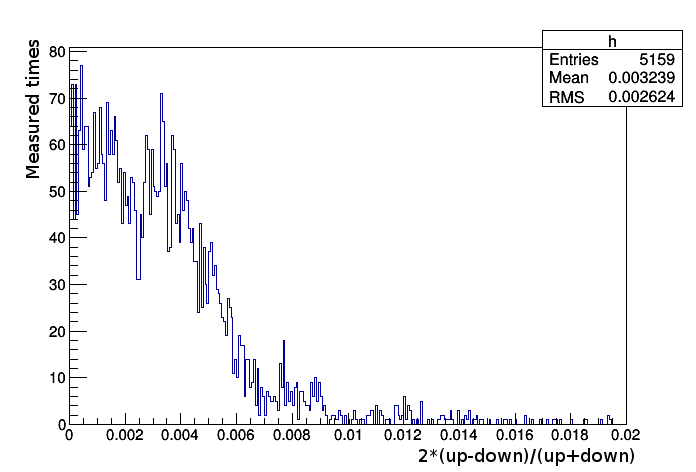}
\par\end{centering}

}
\par\end{centering}

\caption{\label{fig:Comparingud}Comparison of the charge calculated from the upstream and downstream BCMs. Each entry in the graph is the total charge calculated from each run from the experiment.}
\end{figure}

By comparing the difference between the upstream and downstream BCMs, the fluctuations between the two are below $\mathrm{0.19\ \mu C}$ for 90 \% of the runs. The relative differences between them for 90 \% of the runs are below 0.7 \%, as shown in Fig. \ref{fig:Comparingud}
. The differences indicate the uncertainty due to the stability of the BCMs is $\sim$ 0.7 \%. Combined with the calibration uncertainty of the tungsten calorimeter, the total uncertainty of BCMs is $\sim$ 1 \%.

\begin{figure}
\begin{centering}
\includegraphics[width=0.8\linewidth]{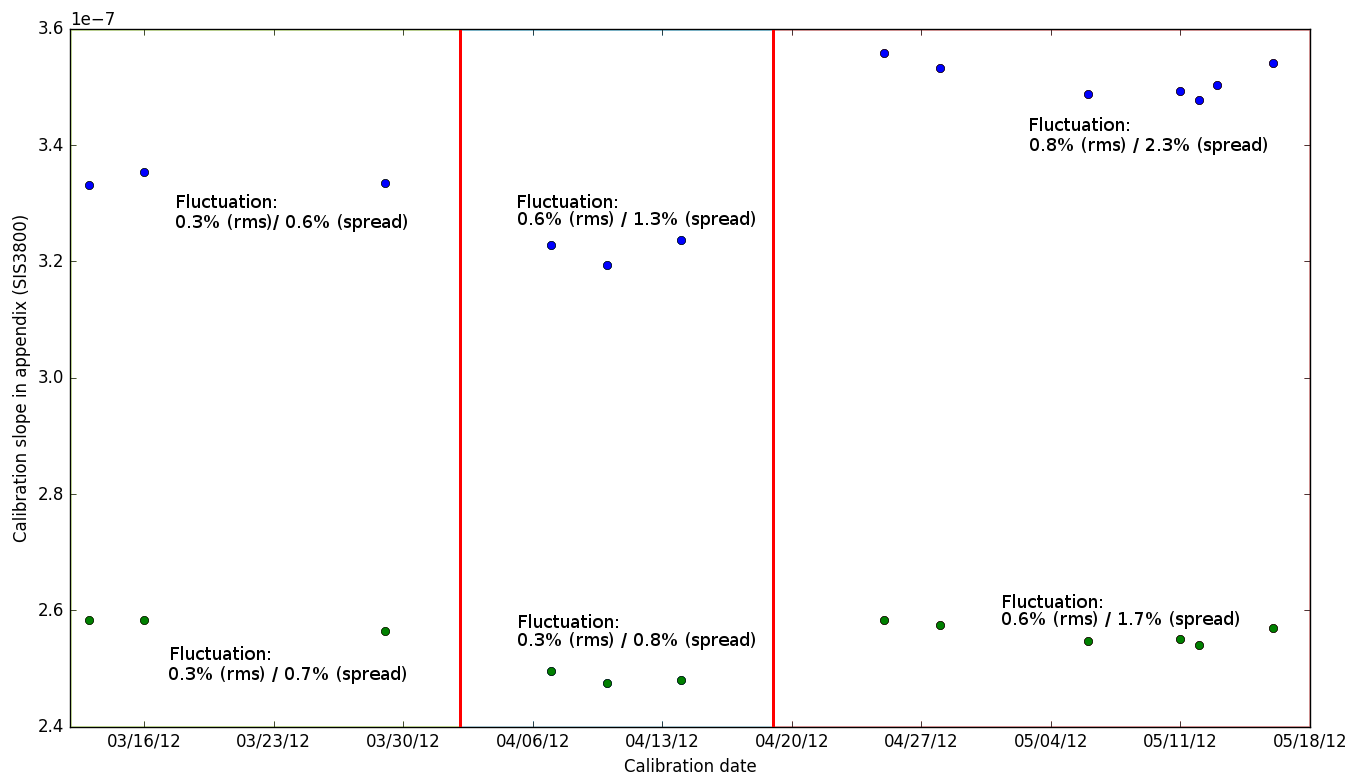}
\par\end{centering}

\caption{\label{fig:BCM-timestab}BCM calibration constants change during 3/13/2012 - 5/18/2012. The blue dot is for upstream BCM, and the green dot is for downstream BCM (recorded in the SIS3800 scaler in right arm). The x axis is the calibration date, while the y axis is the slope recorded in the Appendix (SIS3800). The constants were changed in Apr.2 and Apr.19 (splitted with red line). The fluctuations are calculated using root mean square (number on left) and $2\cdot(max-min)/(max+min)$ (number on right), with the date range of Mar.3 $\sim$ Apr.2, Apr.2 $\sim$ Apr.19, and Apr.19 $\sim$ May.18. }
\end{figure}

Fig.  \ref{fig:BCM-timestab} shows the stability of the calibration with time during 3/13/2012 - 5/18/2012. 
 The calibration constants were stable at the level of $\sim$ 1 \% for most of the time with following exceptions when there were condition changes:
\begin{itemize}
\item Begin - Mar.17, third arm downstream scaler abnormal
\item Mar.18 - Apr.2, left arm upstream scaler noisy
\item Apr. 2, calibration constants changed 3$\sim$4\% for both BCMs
\item Apr.2 - Apr.9, right arm SIS3801 not working
\item Apr.9, changed right arm scaler channel for BCM
\item Apr. 19, calibration constants changed $\sim$ 9\% for the upstream BCM and $\sim$ 4\% for the downstream
\item Near May.12, third arm SIS3801 not working
\item May.13 - May.14, downstream BCM broken
\end{itemize}

\section{\label{sec:Calibration-constants}Calibration constants}

The calibration constants are shown in the Appendix Tables 1-9. The gain settings of the bcm receivers for each periods are list below, the values after the date are: 

A\_Pre\_Gain\_1/A\_Pre\_Gain\_2/A\_Mag\_Div

B\_Pre\_Gain\_1/B\_Pre\_Gain\_2/B\_Mag\_Div

IQ\_Filter\_K

Mag\_Filter\_K
\begin{itemize}
\item Begin - Mar.2 18:39:42 Gain changing
\item Mar.2 18:39:43 - Mar.5 10:00:12 10/10/1 9/9/1 3 4
\item Mar.5 10:00:13 - Mar.5 10:27:40 Gain changing 
\item Mar.5 10:27:41 - Mar.6 8:51:39 10/10/1 9/9/1 1 4
\item Mar.6 8:51:40 - Mar.6 13:45:43 10/13/1 9/9/1 1 4
\item Mar.6 13:45:44 - Mar.6 13:49:05 Gain changing
\item Mar.6 13:49:06 - Mar.7 17:20:10 29/30/4 27/27/4 1 4
\item Mar.7 17:20:11 - Mar.7 17:23:25 Gain changing
\item Mar.7 17:23:25 - Mar.10 13:12:53 29/30/4 27/27/4 1 4
\item Mar.10 13:12:54 - Mar.10 13:33:15 Gain changing
\item Mar.10 13:33:16 - end 40/41/4 40/43/4 1 4 
\end{itemize}
\bibliography{bcm_technote}

\section*{Appendix}

\begin{landscape} \begin{table} \begin{tabular}{|c|c|c|c|c|c|c|} \hline current(nA) & 280 & 25 & 50 & 75 & 100\tabularnewline \hline energy(MeV) & 2253.13 & 2252.94 & 2252.94 & 2252.94 & 2252.94\tabularnewline \hline time & 03/03/12 09:30 PM & 03/13/12 04:00 PM & 03/16/12 10:15 PM & 03/29/12 12:21 AM & 04/07/12 03:00 PM\tabularnewline \hline Avail period & Start~3.10 13:25 & 3.10 13:33~3.17 10:00 & 3.10 13:33~3.17 10:00 & 3.27 21:00~4.2 14:00 & 4.2 18:00~4.9 9:00\tabularnewline \hline run avail(left SIS3800 up) & Start~3051 & 3052~3295 & 3052~3295 & broken & 3660~4695\tabularnewline \hline run avail(left SIS3800 down) & Start~3051 & 3052~3634 & 3052~3634 & 3052~3634 & 3636~4695\tabularnewline \hline run avail(left SIS3801 up) & Start~3051 & 3052~3295 & 3052~3295 & broken & 3660~4695\tabularnewline \hline run avail(left SIS3801 down) & Start~3051 & 3052~3634 & 3052~3634 & 3052~3634 & 3636~4695\tabularnewline \hline run avail(left HAPPEX up) & Start~3051 &  &  & 3073~3634 & 3636~4695\tabularnewline \hline run avail(left HAPPEX down) & Start~3051 &  &  & 3073~3634 & 3636~4695\tabularnewline \hline runnumber & 2665 & 3149 & 3254 & 3437 & 3856\tabularnewline \hline SIS3800 upslope(slowclk)  & 1.37309e-06 & 3.32775e-07 & 3.34603E-07 & 2.30692e-07 & 3.22991E-07\tabularnewline \hline SIS3800 uppedslope(slowclk) & 5.24344E+00 & 2.04717E+00 & 2.01363E+00 & 2.40785E+00 & 2.01757E+00\tabularnewline \hline SIS3800 downslope(slowclk)  & 1.30711e-06 & 2.58036e-07 & 2.57738E-07 & 2.56379e-07 & 2.49656E-07\tabularnewline \hline SIS3800 downpedslope(slowclk) & 6.91734E+00 & 2.76818E+00 & 2.73600E+00 & 2.58978E+00 & 2.73382E+00\tabularnewline \hline SIS3800 upslope(fstclk)  & 1.37346e-06 & 3.32774e-07 & 3.34604E-07 & 2.30693e-07 & 3.22990E-07\tabularnewline \hline SIS3800 uppedslope(fstclk) & 5.17426E-02 & 2.01725E-02 & 1.98515E-02 & 2.37558E-02 & 1.98845E-02\tabularnewline \hline SIS3800 downslope(fstclk)  & 1.30736e-06 & 2.58035e-07 & 2.57739E-07 & 2.5638e-07 & 2.49655E-07\tabularnewline \hline SIS3800 downpedslope(fstclk) & 6.82245E-02 & 2.72772E-02 & 2.69731E-02 & 2.55486E-02 & 2.69437E-02\tabularnewline \hline SIS3801 upslope(fstclk)  & 1.47161e-06 & 3.65962e-07 & 3.58951e-07 & 2.4832e-07 & 3.46231e-07\tabularnewline \hline SIS3801 uppedslope(fstclk) & 5.15461E-02 & 3.09438E-02 & 1.98588E-02 & 2.37232E-02 & 1.98880E-02\tabularnewline \hline SIS3801 downslope(fstclk)  & 1.40127e-06 & 2.83756e-07 & 2.76493e-07 & 2.7518e-07 & 2.67616e-07\tabularnewline \hline SIS3801 downpedslope(fstclk) & 6.81238E-02 & 4.11454E-02 & 2.69735E-02 & 2.55914E-02 & 2.69484E-02\tabularnewline \hline HAPPEX upslope/875e-6 & not avail &  &  & 9.76018e-07 & 9.44277E-07\tabularnewline \hline HAPPEX uppedslope & not avail &  &  & 7.47561E+02 & 7.64707E+02\tabularnewline \hline HAPPEX downslope/875e-6 & not avail &  &  & 7.81678e-07 & 7.59977E-07\tabularnewline \hline HAPPEX downpedslope & not avail &  &  & 7.06877E+02 & 7.27266E+02\tabularnewline \hline \end{tabular} \protect\caption{BCM calibration constants for the left arm} \end{table} \begin{table} \begin{tabular}{|c|c|c|c|c|c|c|} \hline current(nA) & 50 & 75 & 50 & 25 & 50\tabularnewline \hline energy(MeV) & 1712.19 & 1708.35 & 1156.7 & 1156.7 & 2253.65\tabularnewline \hline time & 04/10/12 08:09 AM & 04/14/12 07:07 PM & 04/25/12 02:38 AM & 04/28/12 10:15 AM & 05/06/12 02:43 PM\tabularnewline \hline Avail period & 4.10 0:00~4.19 8:00 & 4.10 0:00~4.19 8:00 & 4.20 4:00~5.2 8:00 & 4.20 4:00~5.2 8:00 & 5.2 21:00~5.13 1:00\tabularnewline \hline run avail(left SIS3800 up) & 3660~4695 & 3660~4695 & 4698~5440 & 4698~5440 & 5485~6100\tabularnewline \hline run avail(left SIS3800 down) & 3636~4695 & 3636~4695 & 4698~5440 & 4698~5440 & 5485~6043\tabularnewline \hline run avail(left SIS3801 up) & 3660~4695 & 3660~4695 & 4698~5440 & 4698~5440 & 5485~6100\tabularnewline \hline run avail(left SIS3801 down) & 3636~4695 & 3636~4695 & 4698~5440 & 4698~5440 & 5485~6043\tabularnewline \hline run avail(left HAPPEX up) &  &  & 4698~5440 & 4698~5440 & 5485~6100\tabularnewline \hline run avail(left HAPPEX down) &  &  & 4698~5440 & 4698~5440 & 5485~6043\tabularnewline \hline runnumber & 4088 & 4405 & 5015 & 5214 & 5751\tabularnewline \hline SIS3800 upslope(slowclk)  & 3.19668E-07 & 3.23814E-07 & 3.55483E-07 & 3.53225E-07 & 3.48943E-07\tabularnewline \hline SIS3800 uppedslope(slowclk) & 2.02148E+00 & 2.02217E+00 & 1.99755E+00 & 2.03317E+00 & 2.00619E+00\tabularnewline \hline SIS3800 downslope(slowclk)  & 2.47684E-07 & 2.48227E-07 & 2.58163E-07 & 2.57395E-07 & 2.54841E-07\tabularnewline \hline SIS3800 downpedslope(slowclk) & 2.73704E+00 & 2.74181E+00 & 2.71992E+00 & 2.75600E+00 & 2.74454E+00\tabularnewline \hline SIS3800 upslope(fstclk)  & 3.19669E-07 & 3.23815E-07 & 3.55483E-07 & 3.53221E-07 & 3.48942E-07\tabularnewline \hline SIS3800 uppedslope(fstclk) & 1.99247E-02 & 1.99301E-02 & 1.96837E-02 & 2.00331E-02 & 1.97694E-02\tabularnewline \hline SIS3800 downslope(fstclk)  & 2.47685E-07 & 2.48228E-07 & 2.58163E-07 & 2.57392E-07 & 2.54841E-07\tabularnewline \hline SIS3800 downpedslope(fstclk) & 2.69776E-02 & 2.70227E-02 & 2.68033E-02 & 2.71552E-02 & 2.70453E-02\tabularnewline \hline SIS3801 upslope(fstclk)  & 3.42781e-07 & 3.4717e-07 & 3.81166e-07 & 3.78808e-07 & 3.7418e-07\tabularnewline \hline SIS3801 uppedslope(fstclk) & 1.99254E-02 & 1.99350E-02 & 1.97021E-02 & 2.00361E-02 & 1.97749E-02\tabularnewline \hline SIS3801 downslope(fstclk)  & 2.65591e-07 & 2.6613e-07 & 2.76814e-07 & 2.76033e-07 & 2.73271e-07\tabularnewline \hline SIS3801 downpedslope(fstclk) & 2.69789E-02 & 2.70306E-02 & 2.68191E-02 & 2.71594E-02 & 2.70443E-02\tabularnewline \hline HAPPEX upslope/875e-6 &  &  & 1.04042E-06 & 1.03394E-06 & 1.02106E-06\tabularnewline \hline HAPPEX uppedslope &  &  & 7.58219E+02 & 7.74628E+02 & 7.63704E+02\tabularnewline \hline HAPPEX downslope/875e-6 &  &  & 7.86150E-07 & 7.84135E-07 & 7.76177E-07\tabularnewline \hline HAPPEX downpedslope &  &  & 7.25062E+02 & 7.39477E+02 & 7.35583E+02\tabularnewline \hline \end{tabular} \protect\caption{BCM calibration constants for the left arm} \end{table} \begin{table} \begin{tabular}{|c|c|c|c|c|c|} \hline current(nA) & 75 & 100 & 50 & 75\tabularnewline \hline energy(MeV) & 2253.34 & 2253.37 & 2252.94 & 3352.4\tabularnewline \hline time & 05/11/12 06:26 PM & 05/12/12 05:48 PM & 05/13/12 02:59 PM & 05/16/12 11:41 PM\tabularnewline \hline Avail period & 5.2 21:00~5.13 1:00 & 5.2 21:00~5.13 1:00 & 5.13 1:00~5.14 8:00 & 5.14 15:00~end\tabularnewline \hline run avail(left SIS3800 up) & 5485~6100 & 5485~6100 & 5485~6100 & 6101~end\tabularnewline \hline run avail(left SIS3800 down) & 5485~6043 & 5485~6043 & broken & 6101~end(NR)\tabularnewline \hline run avail(left SIS3801 up) & 5485~6100 & 5485~6100 & 5485~6100 & 6101~end\tabularnewline \hline run avail(left SIS3801 down) & 5485~6043 & 5485~6043 & broken & 6101~end(NR)\tabularnewline \hline run avail(left HAPPEX up) & 5485~6100 & 5485~6100 & 5485~6100 & 6101~end\tabularnewline \hline run avail(left HAPPEX down) & 5485~6043 & 5485~6043 & broken & 6101~end(NR)\tabularnewline \hline runnumber & 5986 & 6035 & 6062 & 6174\tabularnewline \hline SIS3800 upslope(slowclk)  & 3.49032E-07 & 3.47590E-07 & 3.50492E-07 & 3.55051e-07\tabularnewline \hline SIS3800 uppedslope(slowclk) & 2.01442E+00 & 2.01317E+00 & 2.01224E+00 & 2.00844E+00\tabularnewline \hline SIS3800 downslope(slowclk)  & 2.54925E-07 & 2.53945E-07 & 1.96575E-06 & 2.57619e-07\tabularnewline \hline SIS3800 downpedslope(slowclk) & 2.72054E+00 & 2.71869E+00 & 2.71795E+00 & 2.73275E+00\tabularnewline \hline SIS3800 upslope(fstclk)  & 3.49036E-07 & 3.47589E-07 & 3.50490E-07 & 3.55056e-07\tabularnewline \hline SIS3800 uppedslope(fstclk) & 1.98566E-02 & 1.98398E-02 & 1.98255E-02 & 1.97970E-02\tabularnewline \hline SIS3800 downslope(fstclk)  & 2.54928E-07 & 2.53944E-07 & 1.96567E-06 & 2.57623e-07\tabularnewline \hline SIS3800 downpedslope(fstclk) & 2.68156E-02 & 2.67923E-02 & 2.67783E-02 & 2.69365E-02\tabularnewline \hline SIS3801 upslope(fstclk)  & 3.74239e-07 & 3.72691e-07 & 3.75909e-07 & 3.80653e-07\tabularnewline \hline SIS3801 uppedslope(fstclk) & 1.98584E-02 & 1.98459E-02 & 1.98381E-02 & 1.97998E-02\tabularnewline \hline SIS3801 downslope(fstclk)  & 2.73332e-07 & 2.72283e-07 & 2.10829e-06 & 2.76198e-07\tabularnewline \hline SIS3801 downpedslope(fstclk) & 2.68168E-02 & 2.67972E-02 & 2.68024E-02 & 2.69430E-02\tabularnewline \hline HAPPEX upslope/875e-6 & 1.02102E-06 & 1.01675E-06 & 1.02575E-06 & 1.03841e-06\tabularnewline \hline HAPPEX uppedslope & 7.66377E+02 & 7.66642E+02 & 7.65561E+02 & 7.60940E+02\tabularnewline \hline HAPPEX downslope/875e-6 & 7.76257E-07 & 7.73150E-07 & 5.97952E-06 & 7.84284e-07\tabularnewline \hline HAPPEX downpedslope & 7.26879E+02 & 7.26759E+02 & 7.23959E+02 & 7.29082E+02\tabularnewline \hline \end{tabular} \protect\caption{BCM calibration constants for the left arm} \end{table} \begin{table} \begin{tabular}{|c|c|c|c|c|c|c|} \hline current(nA) & 280 & 25 & 50 & 75 & 100\tabularnewline \hline energy(MeV) & 2253.13 & 2252.94 & 2252.94 & 2252.94 & 2252.94\tabularnewline \hline time & 03/03/12 09:30 PM & 03/13/12 04:00 PM & 03/16/12 10:15 PM & 03/29/12 12:21 AM & 04/07/12 03:00 PM\tabularnewline \hline Avail period & Start~3.10 13:25 & 3.10 13:33~3.17 10:00 & 3.10 13:33~3.17 10:00 & 3.27 21:00~4.2 14:00 & 4.2 18:00~4.9 9:00\tabularnewline \hline run avail(right SIS3800 up) & Start~22130 & 22131~22658 & 22131~22658 & 22131~22658 & 22660~22987\tabularnewline \hline run avail(right SIS3800 down) & Start~22130 & 22131~22658 & 22131~22658 & 22131~22658 & 22660~22987\tabularnewline \hline run avail(right SIS3801 up) & Start~22130 & 22131~22658 & 22131~22658 & 22131~22658 & broken\tabularnewline \hline run avail(right SIS3801 down) & Start~22130 & 22131~22658 & 22131~22658 & 22131~22658 & broken\tabularnewline \hline run avail(right HAPPEX up) & Start~22130 &  &  & 22158~22658 & 22660~23618\tabularnewline \hline run avail(right HAPPEX down) & Start~22130 &  &  & 22158~22658 & 22660~23618\tabularnewline \hline runnumber & 21751 & 22238 & 22338 & 22470 & 22885\tabularnewline \hline SIS3800 upslope(slowclk)  & 1.37212e-06 & 3.33141E-07 & 3.35299E-07 & 3.33446E-07 & 3.22819E-07\tabularnewline \hline SIS3800 uppedslope(slowclk) & 5.23480E+00 & 2.04715E+00 & 2.01363E+00 & 1.88817E+00 & 2.01757E+00\tabularnewline \hline SIS3800 downslope(slowclk)  & 1.30632e-06 & 2.58320E-07 & 2.58274E-07 & 2.56520E-07 & 2.49523E-07\tabularnewline \hline SIS3800 downpedslope(slowclk) & 6.91143E+00 & 2.76815E+00 & 2.73599E+00 & 2.58975E+00 & 2.73382E+00\tabularnewline \hline SIS3800 upslope(fstclk)  & 1.37253e-06 & 3.33139E-07 & 3.35301E-07 & 3.33445E-07 & 3.22818E-07\tabularnewline \hline SIS3800 uppedslope(fstclk) & 5.16652E-02 & 2.01720E-02 & 1.98516E-02 & 1.86263E-02 & 1.98844E-02\tabularnewline \hline SIS3800 downslope(fstclk)  & 1.30656e-06 & 2.58318E-07 & 2.58276E-07 & 2.56521E-07 & 2.49521E-07\tabularnewline \hline SIS3800 downpedslope(fstclk) & 6.81632E-02 & 2.72768E-02 & 2.69733E-02 & 2.55485E-02 & 2.69437E-02\tabularnewline \hline SIS3801 upslope(fstclk)  & 1.46934e-06 & 3.65342e-07 & 3.59699e-07 & 3.57709e-07 & broken\tabularnewline \hline SIS3801 uppedslope(fstclk) & 5.16033E-02 & 2.97348E-02 & 1.98585E-02 & 1.86580E-02 & broken\tabularnewline \hline SIS3801 downslope(fstclk)  & 1.39902e-06 & 2.83277e-07 & 2.77067e-07 & 2.75175e-07 & broken\tabularnewline \hline SIS3801 downpedslope(fstclk) & 6.81575E-02 & 3.95893E-02 & 2.69721E-02 & 2.55542E-02 & broken\tabularnewline \hline HAPPEX upslope/875e-6 & not avail &  &  & 9.55044e-07 & 9.23204E-07\tabularnewline \hline HAPPEX uppedslope & not avail &  &  & 1.75740E+03 & 1.75756E+03\tabularnewline \hline HAPPEX downslope/875e-6 & not avail &  &  & 7.81495e-07 & 7.59968E-07\tabularnewline \hline HAPPEX downpedslope & not avail &  &  & 1.07545E+03 & 1.07213E+03\tabularnewline \hline \end{tabular} \protect\caption{BCM calibration constants for the right arm} \end{table} \begin{table} \begin{tabular}{|c|c|c|c|c|c|c|} \hline current(nA) & 50 & 75 & 50 & 25 & 50\tabularnewline \hline energy(MeV) & 1712.19 & 1708.35 & 1156.7 & 1156.7 & 2253.65\tabularnewline \hline time & 04/10/12 08:09 AM & 04/14/12 07:07 PM & 04/25/12 02:38 AM & 04/28/12 10:15 AM & 05/06/12 02:43 PM\tabularnewline \hline Avail period & 4.10 0:00~4.19 8:00 & 4.10 0:00~4.19 8:00 & 4.20 4:00~5.2 8:00 & 4.20 4:00~5.2 8:00 & 5.2 21:00~5.13 1:00\tabularnewline \hline run avail(right SIS3800 up) & 22600~23618 & 22600~23618 & 23621~24216 & 23621~24216 & 24259~24727\tabularnewline \hline run avail(right SIS3800 down) & 22600~23618 & 22600~23618 & 23621~24216 & 23621~24216 & 24259~24706\tabularnewline \hline run avail(right SIS3801 up) & 23075~23618 & 23075~23618 & 23621~24216 & 23621~24216 & 24259~24727\tabularnewline \hline run avail(right SIS3801 down) & 23075~23618 & 23075~23618 & 23621~24216 & 23621~24216 & 24259~24706\tabularnewline \hline run avail(right HAPPEX up) &  &  & 23621~24216 & 23621~24216 & 24259~24727\tabularnewline \hline run avail(right HAPPEX down) &  &  & 23621~24216 & 23621~24216 & 24259~24706\tabularnewline \hline runnumber & 23082 & 23360 & 23890 & 24040 & 24458\tabularnewline \hline SIS3800 upslope(slowclk)  & 3.19449E-07 & 3.23652E-07 & 3.55750E-07 & 3.53327E-07 & 3.48757E-07\tabularnewline \hline SIS3800 uppedslope(slowclk) & 2.02144E+00 & 2.02219E+00 & 1.99750E+00 & 2.03317E+00 & 2.00620E+00\tabularnewline \hline SIS3800 downslope(slowclk)  & 2.47514E-07 & 2.48103E-07 & 2.58357E-07 & 2.57469E-07 & 2.54705E-07\tabularnewline \hline SIS3800 downpedslope(slowclk) & 2.73699E+00 & 2.74179E+00 & 2.71988E+00 & 2.75600E+00 & 2.74454E+00\tabularnewline \hline SIS3800 upslope(fstclk)  & 3.19451E-07 & 3.23653E-07 & 3.55750E-07 & 3.53323E-07 & 3.48756E-07\tabularnewline \hline SIS3800 uppedslope(fstclk) & 1.99248E-02 & 1.99306E-02 & 1.96839E-02 & 2.00331E-02 & 1.97694E-02\tabularnewline \hline SIS3800 downslope(fstclk)  & 2.47516E-07 & 2.48103E-07 & 2.58357E-07 & 2.57466E-07 & 2.54705E-07\tabularnewline \hline SIS3800 downpedslope(fstclk) & 2.69778E-02 & 2.70229E-02 & 2.68034E-02 & 2.71552E-02 & 2.70453E-02\tabularnewline \hline SIS3801 upslope(fstclk)  & 3.42548e-07 & 3.46997e-07 & 3.81454e-07 & 3.78918e-07 & 3.7398e-07\tabularnewline \hline SIS3801 uppedslope(fstclk) & 1.99254E-02 & 1.99350E-02 & 1.97020E-02 & 2.00361E-02 & 1.97749E-02\tabularnewline \hline SIS3801 downslope(fstclk)  & 2.6541e-07 & 2.65997e-07 & 2.77022e-07 & 2.76114e-07 & 2.73125e-07\tabularnewline \hline SIS3801 downpedslope(fstclk) & 2.69789E-02 & 2.70307E-02 & 2.68188E-02 & 2.71595E-02 & 2.70442E-02\tabularnewline \hline HAPPEX upslope/875e-6 &  &  & 1.01836E-06 & 1.01194E-06 & 9.98575E-07\tabularnewline \hline HAPPEX uppedslope &  &  & 1.76323E+03 & 1.77658E+03 & 1.76456E+03\tabularnewline \hline HAPPEX downslope/875e-6 &  &  & 7.86595E-07 & 7.84236E-07 & 7.75621E-07\tabularnewline \hline HAPPEX downpedslope &  &  & 1.09454E+03 & 1.10431E+03 & 1.10078E+03\tabularnewline \hline \end{tabular} \protect\caption{BCM calibration constants for the right arm} \end{table} \begin{table} \begin{tabular}{|c|c|c|c|c|c|} \hline current(nA) & 75 & 100 & 50 & 75\tabularnewline \hline energy(MeV) & 2253.34 & 2253.37 & 2252.94 & 3352.4\tabularnewline \hline time & 05/11/12 06:26 PM & 05/12/12 05:48 PM & 05/13/12 02:59 PM & 05/16/12 11:41 PM\tabularnewline \hline Avail period & 5.2 21:00~5.13 1:00 & 5.2 21:00~5.13 1:00 & 5.13 1:00~5.14 8:00 & 5.14 15:00~end\tabularnewline \hline run avail(right SIS3800 up) & 24259~24727 & 24259~24727 & 24259~24727 & 24728~end\tabularnewline \hline run avail(right SIS3800 down) & 24259~24706 & 24259~24706 & broken & 24728~end(NR)\tabularnewline \hline run avail(right SIS3801 up) & 24259~24727 & 24259~24727 & 24259~24727 & 24728~end\tabularnewline \hline run avail(right SIS3801 down) & 24259~24706 & 24259~24706 & broken & 24728~end(NR)\tabularnewline \hline run avail(right HAPPEX up) & 24259~24727 & 24259~24727 & 24259~24727 & 24728~end\tabularnewline \hline run avail(right HAPPEX down) & 24259~24706 & 24259~24706 & broken & 24728~end(NR)\tabularnewline \hline runnumber & 24671 & 24700 & 24719 & 24769\tabularnewline \hline SIS3800 upslope(slowclk)  & 3.49296E-07 & 3.47708E-07 & 3.50342E-07 & 3.54078E-07\tabularnewline \hline SIS3800 uppedslope(slowclk) & 2.01441E+00 & 2.01317E+00 & 2.01224E+00 & 2.00686E+00\tabularnewline \hline SIS3800 downslope(slowclk)  & 2.55118E-07 & 2.54031E-07 & 1.96491E-06 & 2.56913E-07\tabularnewline \hline SIS3800 downpedslope(slowclk) & 2.72053E+00 & 2.71870E+00 & 2.71795E+00 & 2.73056E+00\tabularnewline \hline SIS3800 upslope(fstclk)  & 3.49301E-07 & 3.47708E-07 & 3.50340E-07 & 3.54078E-07\tabularnewline \hline SIS3800 uppedslope(fstclk) & 1.98572E-02 & 1.98398E-02 & 1.98256E-02 & 1.97767E-02\tabularnewline \hline SIS3800 downslope(fstclk)  & 2.55121E-07 & 2.54031E-07 & 1.96483E-06 & 2.56913E-07\tabularnewline \hline SIS3800 downpedslope(fstclk) & 2.68159E-02 & 2.67923E-02 & 2.67784E-02 & 2.69082E-02\tabularnewline \hline SIS3801 upslope(fstclk)  & 3.74523e-07 & 3.72818e-07 & 3.75748e-07 & 3.79606e-07\tabularnewline \hline SIS3801 uppedslope(fstclk) & 1.98591E-02 & 1.98461E-02 & 1.98382E-02 & 1.97795E-02\tabularnewline \hline SIS3801 downslope(fstclk)  & 2.7354e-07 & 2.72376e-07 & 2.10737e-06 & 2.75437e-07\tabularnewline \hline SIS3801 downpedslope(fstclk) & 2.68172E-02 & 2.67970E-02 & 2.68023E-02 & 2.69125E-02\tabularnewline \hline HAPPEX upslope/875e-6 & 9.99842E-07 & 9.95083E-07 & 1.00323E-06 & 1.01328E-06\tabularnewline \hline HAPPEX uppedslope & 1.77152E+03 & 1.77303E+03 & 1.77264E+03 & 1.76628E+03\tabularnewline \hline HAPPEX downslope/875e-6 & 7.76667E-07 & 7.73271E-07 & 5.97638E-06 & 7.82040E-07\tabularnewline \hline HAPPEX downpedslope & 1.08914E+03 & 1.08845E+03 & 1.08867E+03 & 1.09584E+03\tabularnewline \hline \end{tabular} \protect\caption{BCM calibration constants for the right arm} \end{table} \begin{table} \begin{tabular}{|c|c|c|c|c|c|c|} \hline current(nA) & 280 & 25 & 50 & 75 & 100\tabularnewline \hline energy(MeV) & 2253.13 & 2252.94 & 2252.94 & 2252.94 & 2252.94\tabularnewline \hline time & 03/03/12 09:30 PM & 03/13/12 04:00 PM & 03/16/12 10:15 PM & 03/29/12 12:21 AM & 04/07/12 03:00 PM\tabularnewline \hline Avail period & Start~3.10 13:25 & 3.10 13:33~3.17 10:00 & 3.10 13:33~3.17 10:00 & 3.27 21:00~4.2 14:00 & 4.2 18:00~4.9 9:00\tabularnewline \hline run avail(third SIS3800 up) & not avail & 40296~40668 & 40296~40668 & 40296~40668 & 40670~41419\tabularnewline \hline run avail(third SIS3800 down) & not avail & broken & broken & 40465~40668 & 40670~41419\tabularnewline \hline run avail(third SIS3801 up) & not avail & 40296~40668 & 40296~40668 & 40296~40668 & 40670~41419\tabularnewline \hline run avail(third SIS3801 down) & not avail & broken & broken & 40465~40668 & 40670~41419\tabularnewline \hline runnumber &  & 40368 & 40388 & 40486 & 40928\tabularnewline \hline SIS3800 upslope(fstclk)  & not avail & 3.32885e-07 & 3.34957E-07 & 3.36286e-07 & 3.23371E-07\tabularnewline \hline SIS3800 uppedslope(fstclk) & not avail & 2.01732E-02 & 1.98517E-02 & 1.98266E-02 & 1.98850E-02\tabularnewline \hline SIS3800 downslope(fstclk)  & not avail & 1.2906e-07 & 1.29005E-07 & 2.5868e-07 & 2.49949E-07\tabularnewline \hline SIS3800 downpedslope(fstclk) & not avail & 5.45561E-02 & 5.39468E-02 & 2.68700E-02 & 2.69434E-02\tabularnewline \hline SIS3801 upslope(fstclk)  & not avail & 3.57236e-07 & 3.59326e-07 & 3.60744e-07 & 3.46631e-07\tabularnewline \hline SIS3801 uppedslope(fstclk) & not avail & 2.02169E-02 & 1.98582E-02 & 1.98876E-02 & 1.98876E-02\tabularnewline \hline SIS3801 downslope(fstclk)  & not avail & 1.38495e-07 & 1.38383e-07 & 2.77492e-07 & 2.67924e-07\tabularnewline \hline SIS3801 downpedslope(fstclk) & not avail & 5.46740E-02 & 5.39542E-02 & 2.69487E-02 & 2.69487E-02\tabularnewline \hline \end{tabular} \protect\caption{BCM calibration constants for the third arm} \end{table} \begin{table} \begin{tabular}{|c|c|c|c|c|c|c|} \hline current(nA) & 50 & 75 & 50 & 25 & 50\tabularnewline \hline energy(MeV) & 1712.19 & 1708.35 & 1156.7 & 1156.7 & 2253.65\tabularnewline \hline time & 04/10/12 08:09 AM & 04/14/12 07:07 PM & 04/25/12 02:38 AM & 04/28/12 10:15 AM & 05/06/12 02:43 PM\tabularnewline \hline Avail period & 4.10 0:00~4.19 8:00 & 4.10 0:00~4.19 8:00 & 4.20 4:00~5.2 8:00 & 4.20 4:00~5.2 8:00 & 5.2 21:00~5.13 1:00\tabularnewline \hline run avail(third SIS3800 up) & 40670~41419 & 40670~41419 & 41420~41915 & 41420~41915 & 41922~42052\tabularnewline \hline run avail(third SIS3800 down) & 40670~41419 & 40670~41419 & 41420~41915 & 41420~41915 & 41922~42017\tabularnewline \hline run avail(third SIS3801 up) & 40670~41419 & 40670~41419 & 41420~41915 & 41420~41915 & 41922~42052\tabularnewline \hline run avail(third SIS3801 down) & 40670~41419 & 40670~41419 & 41420~41915 & 41420~41915 & 41922~42017\tabularnewline \hline runnumber & 41027 & 41256 & 41671 & 41846 & 41918\tabularnewline \hline SIS3800 upslope(fstclk)  & 3.19302E-07 & 3.23944E-07 & 3.56080E-07 & 3.53600E-07 & 3.48498E-07\tabularnewline \hline SIS3800 uppedslope(fstclk) & 1.99247E-02 & 1.99308E-02 & 1.96874E-02 & 2.00331E-02 & 1.97662E-02\tabularnewline \hline SIS3800 downslope(fstclk)  & 2.47400E-07 & 2.48326E-07 & 2.58596E-07 & 2.57668E-07 & 2.54516E-07\tabularnewline \hline SIS3800 downpedslope(fstclk) & 2.69776E-02 & 2.70230E-02 & 2.68065E-02 & 2.71551E-02 & 2.70411E-02\tabularnewline \hline SIS3801 upslope(fstclk)  & 3.46631e-07 & 3.47304e-07 & 3.81801e-07 & 3.79211e-07 & 3.73698e-07\tabularnewline \hline SIS3801 uppedslope(fstclk) & 1.98876E-02 & 1.99356E-02 & 1.97044E-02 & 2.00362E-02 & 1.97719E-02\tabularnewline \hline SIS3801 downslope(fstclk)  & 2.67924e-07 & 2.66232e-07 & 2.77274e-07 & 2.76326e-07 & 2.72918e-07\tabularnewline \hline SIS3801 downpedslope(fstclk) & 2.69487E-02 & 2.70302E-02 & 2.68213E-02 & 2.71590E-02 & 2.70400E-02\tabularnewline \hline \end{tabular} \protect\caption{BCM calibration constants for the third arm} \end{table} \begin{table} \begin{tabular}{|c|c|c|c|c|c|} \hline current(nA) & 75 & 100 & 50 & 75\tabularnewline \hline energy(MeV) & 2253.34 & 2253.37 & 2252.94 & 3352.4\tabularnewline \hline time & 05/11/12 06:26 PM & 05/12/12 05:48 PM & 05/13/12 02:59 PM & 05/16/12 11:41 PM\tabularnewline \hline Avail period & 5.2 21:00~5.13 1:00 & 5.2 21:00~5.13 1:00 & 5.13 1:00~5.14 8:00 & 5.14 15:00~end\tabularnewline \hline run avail(third SIS3800 up) & 41922~42052 & 41922~42052 & 41922~42052 & 42053~end\tabularnewline \hline run avail(third SIS3800 down) & 41922~42017 & 41922~42017 & broken & 42053~end(NR)\tabularnewline \hline run avail(third SIS3801 up) & 41922~42052 & not avail & 41922~42052 & 42053~end\tabularnewline \hline run avail(third SIS3801 down) & 41922~42017 & not avail & broken & 42053~end(NR)\tabularnewline \hline runnumber & 41968 & 42008 & 42036 & 42126\tabularnewline \hline SIS3800 upslope(fstclk)  & 3.49381E-07 & 3.47768E-07 & 3.50458E-07 & 3.54554E-07\tabularnewline \hline SIS3800 uppedslope(fstclk) & 1.98550E-02 & 1.98381E-02 & 1.98268E-02 & 1.97758E-02\tabularnewline \hline SIS3800 downslope(fstclk)  & 2.55180E-07 & 2.54075E-07 & 1.96554E-06 & 2.57258E-07\tabularnewline \hline SIS3800 downpedslope(fstclk) & 2.68130E-02 & 2.67914E-02 & 2.67804E-02 & 2.69069E-02\tabularnewline \hline SIS3801 upslope(fstclk)  & 3.74607e-07 & not avail & 3.75868e-07 & 3.80113e-07\tabularnewline \hline SIS3801 uppedslope(fstclk) & 1.98589E-02 & not avail & 1.98389E-02 & 1.97786E-02\tabularnewline \hline SIS3801 downslope(fstclk)  & 2.736e-07 & not avail & 2.10805e-06 & 2.75803e-07\tabularnewline \hline SIS3801 downpedslope(fstclk) & 2.68173E-02 & not avail & 2.68024E-02 & 2.69116E-02\tabularnewline \hline \end{tabular} \protect\caption{BCM calibration constants for the third arm} \end{table} \end{landscape} 
\end{document}